\documentclass[12pt]{revtex4}

\usepackage{graphicx}

\begin{document}

\title[]{On a proposal of superluminal communication}

\author{GianCarlo Ghirardi}
\email{ghirardi@ictp.it}
\affiliation{Department of Physics, University of Trieste, the Abdus Salam ICTP, Trieste  \\
Strada Costiera 11, I-34151 Trieste, Italy}

\author{Raffaele Romano}
\email{rromano@ts.infn.it}
\affiliation{Department of Physics, University of Trieste, Fondazione Parisi, Rome, Italy}

\begin{abstract}
Recently, various new proposals of superluminal transmission of information have appeared in the literature. Since they make systematic resort to recent formal and practical improvements in quantum mechanics, the old theorems proving the impossibility of such a performance must be adapted to the new scenario. In this paper we consider some of the most challenging proposals of this kind and we show why they cannot work.
\end{abstract}



\maketitle


\section{Introduction} Some years ago, a proposal of superluminal communication has been presented by D. Greenberger \cite{Greenberger}. His paper, in spite of its revolutionary character, has not been the object of the detailed investigations which it surely deserves, and at least part of the scientific community considers  the proposal as viable. A quite recent paper \cite{Kalamidas} takes the suggestion by Greenberger  as the motivation for the elaboration of another, in the opinion of the author more feasible, scheme allowing superluminal signaling. Actually, the author of \cite{Kalamidas} states explicitly that the proposal of \cite{Greenberger} ``{\it has not yet been refuted} and calls into question the universality of the no-signaling theorem" (The emphasis is by the author). If this statement would turn out to be true, a general reconsideration of the problem of relativistic causality would be necessary. For this reason it seems useful to spend some time to reconsider and critically discuss the arguments of \cite{Greenberger} and of other related papers \cite{Kalamidas,Jensen1,Jensen2}, which is what we do here.

\section{The proposal by Greenberger} We summarize the procedure of \cite{Greenberger} by focusing our attention on its really crucial points. The physical process  is initiated by a  parametric down conversion in a crystal in which two photons emerge simultaneously by a source S. The two photons are emitted in two different  opposite directions, $(a,a')$ or $(b,b')$, so that their state is the entangled state:

\begin{equation}
\vert \psi \rangle_{1,2} = \frac{1}{\sqrt{2}}[\vert a\rangle_{1}\vert a'\rangle_{2}+\vert b\rangle_{1}\vert b'\rangle_{2}].
\end{equation}

\noindent Subsequently, the two photons  impinge on a series of beam splitters, as shown in figure 1 of \cite{Greenberger}, which we reproduce here to simplify the presentation. The horizontal lines represent the beam splitters which are assumed to both reflect and transmit half the incident light, and produce a phase shift of $\pi/2$ upon reflection and none upon transmission. On the path of the photon emitted along $b$, after it goes through the first beam splitter, there is a phase-shifter A that shifts the phase of any photon passing through it by $\pi$, and that can be inserted or removed from the beam at will.

\begin{figure}[t]
\centering
\begin{center}
 \includegraphics[width=8cm]{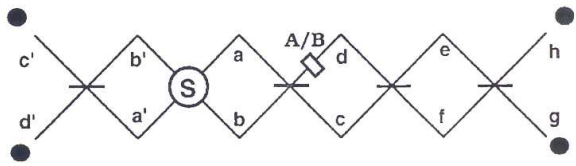} \\
 \caption{\footnotesize Illustration of Greenberger's proposal as depicted in his paper.}\label{fig2}
\end{center}
\end{figure}

At this point the first crucial assumption  of the paper enters into play:

\begin{quote}
 i). {\it The phase shifter} [i.e. a macroscopic object] {\it can be prepared not only in the states $\vert A\rangle$ and $\vert B\rangle$, corresponding to its being inserted or removed from the path of the photon, but also in their orthogonal linear combinations:}
\begin{equation}
\vert u\rangle_3 = \frac{1}{\sqrt{2}}[\vert A\rangle + \vert B\rangle], \qquad
\vert v\rangle_3 = \frac{1}{\sqrt{2}}[\vert A\rangle - \vert B\rangle].
\end{equation}
\end{quote}
According to the author of \cite {Greenberger}, one can also switch on the Hamiltonian $H$,
whose eigenstates are $\vert u \rangle_3$ and $\vert v \rangle_3$, corresponding to slightly different energies, implying the development in time of relative phases with respect to each other.

We will not go through the subsequent elementary calculations of the paper; we limit ourselves to mention that the above assumptions lead to the conclusion that as the photons are nearing their final detectors, represented in the figure by the $4$ black circles, they will be in the following entangled photon-phase shifter state:
\begin{eqnarray}\label{initial}
\vert \psi \rangle_{1,2,3} &=& \frac{1}{2}\big[(-e^{i\alpha}\vert h\rangle_{1}\vert d'\rangle_{2}+e^{-i\alpha}\vert g\rangle_{1}\vert c'\rangle_{2})e^{i\beta}\vert u\rangle_3 \nonumber \\
&+& (e^{i\alpha}\vert g\rangle_{1}\vert c'\rangle_{2}-e^{-i\alpha}\vert h\rangle_{1}\vert d'\rangle_{2})e^{-i\beta}\vert v\rangle_3\big],
\end{eqnarray}
the phase factors $e^{\pm i\alpha}$ and $e^{\pm i\beta}$ being due to the evolution of the states $\vert u\rangle_3$ and $\vert v\rangle_3$ under the effect of $H$.

Now, the author of \cite{Greenberger} puts forward his other crucial assumption. In his words:
\begin{quote}
ii). {\it  In accordance with our assumption that one can manipulate these Cat states, one can turn off $H$ for the state $\vert v\rangle$, while leaving it in place for the state $\vert u\rangle$. This will rotate the state $\vert v\rangle$ into the state $e^{i\gamma}\vert u\rangle$, where $\gamma$ is the accumulated phase difference during this process.}
\end{quote}

As it is obvious and as the author plainly admits, this amounts to accept that a nonunitary transformation $T$ can be performed within the manifold spanned by the two states of the phase shifter, and acts  as follows:
\begin{equation}\label{madtransf}
T\vert u\rangle_3 =\vert u\rangle_3,\qquad T\vert v\rangle_3=e^{i\gamma}\vert u\rangle_3.
\end{equation}
To make plausible this assumption, Greenberger makes explicit reference to an experiment by Mandel \cite{Mandel1,Mandel2} in which a similar nonunitary map plays, in his opinion, an analogous and basic role.

The conclusion follows. After  this transformation the state becomes:
\begin{eqnarray}\label{crucial}
\vert \psi_{final}\rangle_{1,2,3} &=& e^{i\gamma/2} \big[-\cos (\alpha+\beta-\gamma/2)\vert h\rangle_{1}\vert d'\rangle_{2} \nonumber \\
&+& \cos (\beta-\alpha-\gamma/2)\vert g\rangle_{1}\vert c'\rangle_{2}\big]\vert u\rangle_3.
\end{eqnarray}
And now the game is over: by appropriately choosing the angles $\alpha,\beta$ and $\gamma$, one can, at his free will, suppress one of the two terms of the superposition of the photon states, i.e. one can make certain either the firing of the detector in $d'$ or the one in $c'$ (and correspondingly the one in $h$ or the one in $g$) allowing in this way a superluminal transfer of information from the phase shifter, which acts as the signaler, to the photon detectors.

\section{Criticism to the assumptions} Greenberger is perfectly aware that the assumption that one can handle macroscopically different  states of a macroscopic object and, in particular, prepare linear superpositions of states corresponding to different locations for it, is not a trivial one. Actually, as it follows from the analysis of the so called environment induced superselection rules, or, equivalently, by the attempts to solve the macro-objectification problem by resorting to decoherence \footnote{A position that we do not consider appropriate for solving problems of principle, but which has the merit of making  fully clear that the task that Greenberg is trying to solve is impossible in practice.}, preparing the involved superpositions turns out not to be practically feasible.
We do not assign a fundamental role to this serious limitation of the proposal, because we agree with Greenberger that probably one can devise an analogous procedure involving microsystems.

In our opinion, the really unacceptable assumption of the paper is the one we have labeled ii). To discuss this point we begin by stressing that the state (\ref{initial}), which will be subjected to the nonunitary transformation leading to the crucial state (\ref{crucial}), is the linear superposition of the products of two states of a quantum system (the system of the photons) times two orthogonal states of another quantum (according to the author) system, i.e., the phase shifter. Now the crucial assumption summarized by (\ref{madtransf}) is that  a transformation which does not involve the photon states but only the phase shifter and transforms one of its two orthogonal states  into the other one (apart from a phase factor which must be controllable by the experimenter) can be implemented in practice. This fact looks quite peculiar and not fitting within the standard quantum scenario. Moreover, if a transformation of the type devised by Greenberger would be possible, the whole procedure  of Greenberger could be made trivial, avoiding the need of parametric down converters, chains of beam splitters, superpositions of macrostates and so on.

To illustrate this point we consider an elementary EPR-Bohm like setup for two far away spin $1/2$ particles in the singlet state:

\begin{equation}\label{singlet}
\vert \psi_{-}\rangle=\frac{1}{\sqrt{2}}[\vert \uparrow_{1}\rangle \vert \downarrow_{2}\rangle -\vert \downarrow_{1}\rangle \vert \uparrow_{2}\rangle].
\end{equation}

\noindent In strict analogy with what has been assumed by Greenberger, suppose now we can rotate only one of the two spin states of particle $2$ making it to coincide, apart from a controllable phase, with the other one:
\begin{equation}\label{transf}
T\vert \downarrow_{2}\rangle = \vert \downarrow_{2}\rangle, \qquad T \vert \uparrow_{2}\rangle=e^{i\gamma}\vert \downarrow_{2}\rangle.
\end{equation}
Under this transformation the state (\ref{singlet}) becomes a factorized state of the two particles:
\begin{equation}\label{final2}
\vert \psi_{T}\rangle = \frac{1}{\sqrt{2}}[\vert \uparrow_{1}\rangle - e^{i\gamma}\vert \downarrow_{1}\rangle] \vert \downarrow_{2}\rangle
\end{equation}
In (8), the state referring to particle $1$ is an eigenstate of $\sigma\cdot {\bf d}$ for the direction ${\bf d}=(\cos \gamma, \sin \gamma, 0)$ pertaining to the eigenvalue $-1$. This means that a measurement of this observable by Alice (where particle 1 is) will give with certainty the outcome $-1$ if Bob has performed the transformation $T$ on his particle, while, if Bob does nothing, the probability of getting such an outcome equals $1/2$. Having such a device, one can easily implement superluminal transfer of information. Concluding: if assumption ii) were correct, one would not need all the complex apparatus involved in Greenberger's proposal to get the desired result.

Greenberger himself seems aware that the nonunitary transformation $T$ in (\ref{madtransf}) has strict relations with the transformation we have summarized in (\ref{transf}). Actually, when he mentions for the first time this procedure, he states that it does not differ from the one of rotating, by switching on a magnetic field, from spin up to spin down the state of a particle in one arm of an interferometer. This remark offers the possibility of making clear, by resorting to our oversimplified treatment, where the argument of Greenberger fails. In fact, in the spin analogous of his proposal, he needs to rotate {\it only one of the two terms of a superposition and not the whole statevector of a single particle}, which is a quite different story.

To further clarify this point let us consider the singlet state (\ref{singlet}) and let us switch on an inhomogeneous magnetic field along the z-axis on the Bob side of the experiment, i.e., let us subject particle $2$ to a Stern-Gerlach process. In this case the evolution of the spin eigenstates for such a particle can be described in the following manner:
\begin{equation}
\vert \downarrow_{2}\rangle \rightarrow \vert \downarrow_{2}, \Downarrow_{2}\rangle,\;\;\vert \uparrow_{2}\rangle \rightarrow \vert \uparrow_{2}, \Uparrow_{2}\rangle,
\end{equation}
where the states $\vert \Downarrow_{2}\rangle$ and $\vert \Uparrow_{2}\rangle$ correspond to propagation of the particle along a downward or upward space direction, respectively.

\noindent Now, in full agreement with Greenberger's remark, it is possible to switch on a homogeneous magnetic field confined, e.g., to the upward path, such that a rotation of the spin from the upward to the downward spin direction occurs (even with a phase change $e^{i\gamma}$) so that the two final states are essentially the same state and can be factored out as it happens for the state $\vert u\rangle_3$ of the phase shifter in (\ref{crucial}). But what remains is by no means a state like (\ref{final2}), corresponding to a definite outcome for the appropriate measurement of a spin component of particle $1$, but the state:
\begin{equation}\label{final3}
\vert \psi_{SG}\rangle=\frac{1}{\sqrt{2}}[\vert \uparrow_{1}\rangle\vert \Downarrow_{2} \rangle - e^{i\gamma}\vert \downarrow_{1}\rangle\vert \Uparrow_{2}\rangle  ] \vert \downarrow_{2}\rangle.
\end{equation}
This state attaches to spin measurements on particle $1$ precisely the same probabilities as the original singlet state, since the spin states of particle $1$ are still entangled with particle $2$. The only result we have obtained is that of transforming the (spin $1$ - spin $2$) entanglement of  the singlet, in the (spin $1$ - location $2$) entanglement of state (\ref{final3}). Therefore, the just mentioned pseudo-nonunitary (in the spin-space) transformation  cannot be used for superluminal communication. Obviously, one might be tempted to perform an action which transforms the states $\vert \Downarrow_{2}\rangle$ and $\vert \Uparrow_{2}\rangle$ into the same state, i.e. to recombine them. But this would require interactions with further systems and will entangle the spin states of particle 1 with different states of such systems.

One might argue that our analysis, being too simplified, does not properly  describe the operation $T$. Therefore, we pass to analyze in a more detailed way the general case.

\section{Deepening the argument} Suppose one has a composite system whose constituents are denoted as $X$ and $Y$, respectively, the corresponding Hilbert spaces being ${\cal H}_X$ and ${\cal H}_Y$. We assume that the two subsystems are confined in two far away space regions, and that the composite system $X + Y$ and the external world are initially uncorrelated. Moreover, we will consider only local physical actions on one (i.e. $Y$) of the constituents.

We recall that, within the general framework of quantum theory, it is useful to resort to the statistical operator language which is unavoidably brought into the game when general operations are taken into account. From this point of view, the general mathematical maps which correspond to any physical procedure involving only system $Y$  must be completely positive and can be deterministic or probabilistic. Completely positive deterministic transformations can always \cite{Kraus} be written in the form
\begin{equation}\label{krausfinal}
    \bar {\rho}(X,Y) = \sum_{i} (I_X \otimes A_{i}) \rho(X,Y) (I_X \otimes A_{i}^{\dag}),
\end{equation}
where $I_X$ is the identity operator on ${\cal H}_X$, and the so-called {\it Kraus operators} $A_{i}$ are only constrained by
\begin{equation}\label{conskraus}
    \sum_i A^{\dag}_{i}A_{i} = I_Y,
\end{equation}
reflecting their  deterministic character.

Probabilistic operations (e.g., ideal or non-ideal selective measurement processes, with a specified outcome) can be written as
\begin{equation}\label{krausfinal2}
    \bar {\rho}(X,Y) = \frac{1}{p} \sum_{i} (I_X \otimes A_{i}) \rho(X,Y) (I_X \otimes A_{i}^{\dag}),
\end{equation}
where
\begin{equation}\label{conskraus2}
     p = \sum_i {\rm Tr} A_{i}^{\dag}A_{i} \rho(X,Y), \qquad \sum_i A^{\dag}_{i}A_{i} < I_Y.
\end{equation}
Here $p$ is the probability that the operation  actually succeeds. Within quantum mechanics, the paradigmatic example of such a process  is a selective projective ideal measurement of an observable. In such a case one can drop the sum over $i$ and identify  $A_{i}$ with the projector $P_{i}$ on the manifold associated to the considered outcome.

The maps (\ref{krausfinal})  embody all the possible deterministic and probability preserving transformations to which subsystem $Y$ of the the bipartite system can be subjected to. They include  local unitary transformations (deterministic maps with only one Kraus operator), non-selective projective measurements (deterministic maps with Kraus operators given by the eigenprojectors of the measured observable), and non-selective  non-ideal measurements processes in which more general effects replace the projectors of ideal measurements.

%
%
%

On the other hand, the probabilistic nature  of operations of type (\ref{krausfinal2}) implies that they cannot be used for superluminal signaling. For instance,  when one has an entangled system of two far away constituents and the sender at one wing performs a selective measurement, in order that the receiver can take advantage of the fact that reduction took place, he must be informed concerning the actual result obtained by the sender.

Coming back to our argument we consider, for simplicity, the following entangled state:
\begin{equation}\label{psi}
\vert\psi (x,y)\rangle = a_{1} \vert \phi_{1}(x)\rangle \otimes  \vert \chi_{1}(y)\rangle+a_{2} \vert \phi_{2}(x)\rangle \otimes  \vert \chi_{2}(y)\rangle,
\end{equation}
with $\langle\phi_{i}\vert\phi_{j}\rangle= \langle\chi_{i}\vert\chi_{j}\rangle = \delta_{ij}$, and $\vert a_1 \vert^2 + \vert a_2 \vert^2 = 1$, and we formulate a general theorem.

{\it Theorem}: Any transformation of the type of the one considered by Greenberger, which transforms the original state  $\vert\psi(x,y)\rangle $ into $\vert\varphi(x,y)\rangle$, with
\begin{equation}\label{phi}
\vert\varphi(x,y)\rangle = e^{i\eta}[a_{1} \vert \phi_{1}(x)\rangle  + e^{i\gamma}a_{2} \vert \phi_{2}(x)\rangle ]\otimes  \vert \chi_{2}(y)\rangle,
\end{equation}
cannot be described by a completely positive map.

{\it Proof}: Let us concentrate our attention on subsystem $X$ by taking
the partial trace over $Y$ of the right hand side of (\ref{krausfinal}). Taking advantage of the cyclic property of the trace and of (\ref{krausfinal}),
it follows that the reduced statistical operator is left unaffected after any possible action, i.e. :
\begin{equation}
\check{\rho}(X) = {\rm Tr} \bar {\rho}(X,Y) = \vert a_{1}\vert^{2}\vert\phi_{1}\rangle\langle\phi_{1}\vert +\vert a_{2}\vert^{2}\vert\phi_{2}\rangle\langle\phi_{2}\vert.
\end{equation}
On the contrary, for the state (\ref{phi}), which is the one considered by Greenberger, the same operation of partial trace on $Y$ gives the statistical operator:
\begin{equation}
\check{\rho}_T(X) = [a_{1}\vert\phi_{1}(x)\rangle+e^{i\gamma}a_{2}\vert \phi_{2}\rangle] [a_{1}^{*} \langle\phi_{1}(x)\vert+e^{-i\gamma}a_{2}^{*}\langle \phi_{2}\vert],
\end{equation}
corresponding to a pure state for subsystem $X$. As a consequence, Greenberger's transformation cannot correspond to any physically implementable deterministic map\footnote{We have chosen a general approach to derive our theorem just because ref.[1] is based on a peculiar nonunitary map. Alternatively, one might reject Greenberger's proposal by resorting to the old argument \cite{Eber,Gh} that, being the constituents space-like separated, each of them evolves unitarily so that, e.g., Bob's probabilities are obtained by tracing out Alice's system, and thus do not depend on Alice's actions.}. On the other hand, in case of probabilistic operations, taking the partial trace leads to a state for particle $1$ which actually differs from $\check {\rho}(X)$, but, as already remarked, to take advantage of this fact the receiver should already know the outcome obtained by the sender. This shows that the procedure considered by Greenberger cannot be implemented by any of the physically acceptable maps.

\section{Other recent proposals and conclusions} The analysis we have performed for  Greenberg's proposal can be transferred, almost unchanged, to the one by Kalamidas \cite{Kalamidas}, so that we will not spend time to reconsider it. However, a recent different proposal for superluminal communication has been put forward by  R.W. Jensen in two related articles \cite{Jensen1, Jensen2}. In \cite{Jensen2} he claims to be able to overcome the no go theorem  by Eberhard and Ross \cite{Eberhard} concerning superluminal effects, since his procedure is different  from the one considered by these authors due to the fact that he makes resort to what he calls {\it erasing the state}, but is, de facto, a selective measurement.
He states: {\it Physically a particle eigenvalue or path information is erased if we allow the particle to be detected without measuring its eigenvalue information}. This statement amounts to claim  that the effect of erasure on the statistical operator is given by the map corresponding to a nonselective measurement.
On the contrary, when coming the the practical implementation of it, he assumes that erasure amounts to transform the statistical operator subjecting it to the transformation corresponding to a selective measurement. This can obviously be implemented, but it is of no use from a practical point of view just because, as already remarked, the only way to use the process for transmitting information is that the sender informs the receiver about the specific outcome he has got  by sending him a luminal signal.
The conclusion is quite simple: Eberhard and Ross are perfectly right and the proposal under discussion has no physical meaning, in particular, for what concerns us here, it does not allow superluminal signaling.

To provide a consistent model Greenberger should clarify some points:
\begin{enumerate}
\item Is the transformation (5), which is crucial for his proposal, a linear transformation or not? If he takes seriously nonlinear maps, then we must mention that his result is already contained in an old paper by Gisin \cite{Gisin1,Gisin2} who has proved that nonlinearity allows superluminal signaling \footnote{Actually, Gisin has derived his proof to show that Weinberg's \cite{Weinberg1,Weinberg2,Weinberg3} proposal of nonlinear but deterministic modifications of quantum theory violates relativistic causality. On the contrary, modifications of the theory which are both nonlinear and stochastic, as those characterizing collapse theories \cite{Ghirardi, Tumulka, Bedingham}, can be made fully consistent with relativistic requirements.}. But what does suggest to take into account nonlinear transformations within standard quantum mechanics?
\item He should make at least precise the effect on the statistical operator of the physical process he has in mind. As we have already remarked, one must violate the complete positivity of the map in order to hope to attain his aim. But violation of complete positivity is physically very hard to swallow and even more hard to implement by a physical operation.
\item He should be more precise on his statement that he can rotate a spin of one of the two spatial branches of the wave function (by putting a magnetic field only on one of them) because, as we have shown, this does not alter in any way the fact that the physics of system $X$ is not changed by the action on $Y$; simply a new type of entanglement emerges which does not affects the physics of system $X$.
\item He should clarify his claims that Mandel has considered and tested a process of the kind he has in mind. In our opinion, in no step of his experiment Mandel introduces a nonunitary evolution of his system.

\end{enumerate}

Concluding, we have shown that the universality of the no-signaling theorem has not been disproved by the proposals we have analyzed in this paper.

\acknowledgments One of us, R.R., acknowledges the financial support by the R. Parisi Foundation. This research is partially supported by the ARO MURI grant W911NF-11-1-0268.


\section*{References}

\begin{thebibliography}{99}


\bibitem{Greenberger} Greenberger D M 1998 Physica Scripta T {\bf 76} 57

\bibitem{Kalamidas} Kalamidas D A A proposal for a feasible quantum-optical experiment to test the validity of the no-signaling theorem {\it Preprint} arXiv:1110.4629

\bibitem{Eber} Eberhard P 1978 Nuovo Cim. B {\bf 46} 392

\bibitem{Gh} Ghirardi G C, Rimini A and Weber T 1980 Lett. Nuovo Cim. {\bf 27}  293

\bibitem{Jensen1} Jensen R W 2006 AIP Conf. Proc. {\bf 813} 1409

\bibitem{Jensen2} Jensen R W 2010 2009 AIP Conf. Proc. {\bf 1208} 274

\bibitem{Mandel1} Zou X Y, Wang L J and Mandel L 1991 Phys. Rev. Lett. {\bf 67} 318
\bibitem{Mandel2} Wang L J, Zou X Y and Mandel L 1991 Phys. Rev. A {\bf 44} 4614

\bibitem{Kraus} Kraus K 1983 Fundamental Notions of Quantum Theory, Academic Press, Berlin


\bibitem{Eberhard} Eberhard P and Ross R 1989 Found. Phys. Lett. {\bf 2} 127

\bibitem{Gisin1} Gisin N 1989 Helvetica Physica Acta {\bf 62} 363
\bibitem{Gisin2} Gisin N 1990 Phys. Lett. A {\bf 143} 1

\bibitem{Weinberg1} Weinberg S 1989 Phys. Rev. Lett. {\bf 62} 485
\bibitem{Weinberg2} Weinberg S 1989 Ann. Phys. {\bf 194} 336
\bibitem{Weinberg3} Weinberg S 1989 Nuclear Physics B {\bf 6} 67

\bibitem{Ghirardi} Ghirardi G C, Rimini A and Weber T 1986 Phys. Rev. D {\bf 34} 470

\bibitem{Tumulka} Tumulka R 2006 J. Stat. Phys. {\bf 125} 821

\bibitem{Bedingham} Bedingham D, Duerr D, Ghirardi G C, Goldstein S and Zangh\`i N 2011 Matter density and relativistic models of wave function collapse {\it Preprint} arXiv:1111.1425


\end{thebibliography}
\end{document}